\input epsf

\documentstyle[prd,floats,aps,psfig]{revtex}
\tighten 

\newlength{\xyy}
\begin{document}
\draft

\twocolumn[\hsize\textwidth\columnwidth\hsize\csname @twocolumnfalse\endcsname
\preprint{AEI-2000-016}

\title{Making use of geometrical invariants in black hole collisions.}

\author{John Baker and Manuela Campanelli}

\address{
Albert-Einstein-Institut,
Max-Planck-Institut f{\"u}r Gravitationsphysik,
Am M\"uhlenberg 1, D-14476 Golm, Germany}

\date{\today}

\maketitle
\begin{abstract}
We consider curvature invariants in the context of black hole 
collision simulations. In particular, 
we propose a simple and elegant combination of the 
Weyl invariants $I$ and $J$, the {\sl speciality index} 
${\cal S}$.  In the context of black hole perturbations $\cal S$ provides 
a measure of the size of the distortions from 
an ideal Kerr black hole spacetime.
Explicit calculations in well-known examples of axisymmetric 
black hole collisions demonstrate that this quantity may serve 
as a useful tool for predicting in which cases perturbative 
dynamics provide an accurate estimate of the radiation 
waveform and energy.
This makes ${\cal S}$ particularly suited to studying the 
transition from nonlinear to linear dynamics and 
for invariant interpretation of numerical results.

\end{abstract}

\pacs{AEI-2000-016; 04.25.Dm, 04.25.Nx, 04.30.Db, 04.70.Bw}

\vskip2pc]


\section{Introduction}
\label{sec: Introduction}

Einstein's theory of gravity demands the equivalence of all coordinate 
representations of gravitational dynamics.  This coordinate gauge 
invariance makes general relativity very simple and beautiful because 
there are no special families of observers to be considered.  
On the other hand, though, it can also make it difficult to distinguish 
whether differences observed in two spacetime representations are 
true physical (geometrical) differences or gauge differences.  
A natural way to limit confusion between gauge and physical 
differences is to work, whenever possible, with geometrically 
defined scalars, which are invariant under (passive) coordinate 
transformations. Geometric curvature invariants have a long 
productive history in the classification and distinction of exact 
analytic solutions of Einstein's equations, particularly for the 
algebraic Petrov classification and characterization 
of curvature singularities \cite{Kramer80}.
To a lesser extent, curvature invariants have also been applied in 
the field of numerical relativity, primarily for code testing when 
evolving exact solutions numerically. Here gauge invariant methods
have the distinct advantage that they can be applied to evolutions
using numerically generated coordinates which are not understood 
analytically.
What seems to have received less attention is the application of 
curvature invariants in a region of spacetime which is 
perturbatively close to an exact solution of Einstein's equations.  

Black hole perturbation theory has recently generated much interest 
as a model for the late stages of a black hole collision 
spacetime \cite{Price94a}.
When two black holes are close to each other one can simply treat 
the problem as a single distorted black hole that `rings down'
into its final equilibrium state. Perturbative calculations  
applied in this final regime have become an important 
tool in the verification and interpretation of numerically 
generated results \cite{Anninos95g,Baker96a,Baker99a}. More ambitiously, 
the perturbative approach can be used in conjunction with full scale 
3D numerical relativity simulations to directly ``take over'' and 
continue a previously computed numerical  black hole 
spacetime\cite{Baker2000b}. 
It is in the context of such an approach to black hole collisions 
that the authors expect to make special use of curvature invariants.
 
In setting up such a perturbative approach one encounters a family of 
Cauchy data sets which encode distorted black holes such as a family 
of slices from the late-time region of a numerically calculated 
black hole coalescence. We would expect, on physical grounds, that 
for some of the initial data sets, perturbation theory 
can provide an accurate model of the corresponding future-evolved 
spacetime. We would like to have a working criteria for when we 
can expect perturbation theory to be effective based only on numerical 
data.
 
Motivated by this purpose we introduce an invariant quantity ${\cal S}$, 
which geometrically measures local deviation from algebraic 
speciality, and which we expect to be a very useful tool for 
numerical and perturbative work involving near-stationary regions 
of black hole spacetimes.
Other indicators of the potential success of perturbation theory,
 like for instance the size of the distortions of
the apparent horizon, have been previously applied to numerical 
results \cite{Abrahams95d}, but these require 
establishing a coordinate system by an intuitively reasonable 
method and, in most of the cases, are computationally 
expensive and only applicable when perturbations are reasonably 
small. Our invariant index is simple, elegant and has none of 
these shortcomings.   
In fact, it is not at all limited to perturbation studies, but 
can be applied directly to full 3D 
numerical evolutions to directly explore 
the transition from nonlinear to the linear regime\cite{Baker2000b},
or for invariant interpretation
of numerical spacetimes. 
In all well-known examples of axisymmetric 
black hole collisions we studied, ${\cal S}$ has already proven to 
be very useful.

\section{The speciality index}
\label{sec: Invariants}
   
Curvature invariants are part of the standard analysis of exact 
solutions of Einstein's equations.  From the Weyl tensor, $C_{abcd}$, 
which carries information about the gravitational fields in the spacetime, 
one can algebraically derive two complex curvature invariants usually 
called $I$ and $J$.  These are essentially the square and cube the 
self-dual part, 
$\tilde C_{abcd}=C_{abcd}+(i/2)\epsilon_{abmn}C^{mn}_{\,\,\,\,cd}$, of 
the Weyl tensor:
\begin{equation}
I=\tilde C_{abcd}\tilde C^{abcd} ~~{\rm and}~~ 
J=\tilde C_{abcd}\tilde C^{cd}_{\,\,\,\,mn}\tilde C^{mnab}.
\end{equation}
These scalars are useful in the algebraic classification of exact solutions.  
The different algebraic Petrov types are distinguished by the degeneracies 
among the (up to) four principal null directions (PNDs) associated pointwise 
with the Weyl tensor. Type I is the algebraically general case with four 
distinct principal null directions.  The other types II, III, D, and N have at 
least two coincident PNDs and are referred to as algebraically special.  
A notable characteristic common to all stationary isolated 
black hole solutions of general relativity is that they are all 
algebraically special, of Type D, with two pairs of coincident principal 
null directions (PNDs) at each point.
We contend however that for interesting cases involving nontrivial dynamics, 
perturbed black hole spacetimes are generically not algebraically special, of
type I. 

Significantly, the invariants $I$ and $J$ satisfy the relation, 
$I^3=27J^2$ if and only if the Weyl tensor is algebraically special
\cite{Kramer80}. Since the Weyl tensor in a perturbed black hole 
spacetime is not expected to be algebraically special, we expect 
$I^3\neq27J^2$.  
Our proposal, then, is to use this relation to construct an invariant 
index for algebraic speciality as a local measure of the size of 
the distortions from some background black hole. This violation can 
be in general quantified by considering the following 
{\em speciality index}
\begin{equation}
{\cal S} = \frac{27J^2}{I^3}.
\end{equation}
For the unperturbed algebraically special background Kerr spacetime 
${\cal S}=1$.  In the perturbed spacetime we generically expect 
${\cal S}=1+\Delta{\cal S}$, and the size of the deviation 
$\Delta{\cal S}\neq 0$ can be used as a guide to predicting 
the effectiveness of black hole perturbation theory.
  
The theory of perturbations on a background Kerr spacetime was worked out 
first by Teukolsky \cite{Teukolsky73} and has been extensively studied 
by many authors 
\cite{Campanelli99,Campanelli98a,Campanelli98b,Campanelli98c}.  
In this context it is natural to use the Newman-Penrose decomposition 
of the Weyl tensor into five complex quantities, $\psi_0$, $\psi_1$, 
$\psi_2$, $\psi_3$, and $\psi_4$, defined with respect to some choice 
of a null tetrad basis.    
In terms of the Weyl components, for an arbitrary tetrad choice:
\begin{eqnarray}
I&=&3{\psi_2 }^2-4\psi_1\psi_3+\psi_4\psi_0, \nonumber \\
J&=&-\psi_2^3+\psi_0\psi_4\psi_2 + 2\psi_1\psi_3\psi_2-\psi_4\psi_1^2
    -\psi_0\psi_3^2.
\end{eqnarray}
For any Type D spacetime such as Kerr, a tetrad can be conveniently 
chosen such that only $\psi_2$ is non-vanishing.
By expressing ${\cal S}$ with respect to {\it any} perturbation of 
such a tetrad we find that
\begin{equation}
{\cal S}=1-3\epsilon^2\frac{\psi_0^{(1)}\psi_4^{(1)}}
{(\psi_2^{(0)})^2}+{\cal O}(\epsilon^3),
\label{S2}
\end{equation}
were $\epsilon$ is a perturbation parameter, and 
the superscript ${(0)}$ and ${(1)}$ stand respectively for 
background and first order pieces of the perturbed Weyl scalars. Thus, 
the lowest order term in the deviation is second order in the 
perturbation parameter $\epsilon$. In the perturbative context
this means that, when the speciality index is significantly 
different from unity, one can see that a potentially second 
order quantity has become significant, and the first order 
theory should no longer be trusted. 
For the case of Schwarzschild Eq. (\ref{S2}) can be reexpressed 
in terms of the gauge invariant Moncrief functions.\cite{Campanelli98a}.

\section{Example applications}
\label{sec: Black hole}

Since ${\cal S}=1$ in the background, a reasonable 
criterion for when we can expect perturbation theory to be 
applicable is that ${\cal S}$ should differ from one by no 
more than a ``factor of two''.  In regions where this condition 
is violated we can expect significant violations of the 
perturbative dynamics. As test cases, to verify our interpretation 
of ${\cal S}$  we have considered some well-studied cases of 
axisymmetric black hole collisions.  We find it useful to consider 
the location of such violating regions with respect to the 
background black hole horizon and the perturbative 
``potential barrier.''  As is well-known, perturbative black 
hole dynamics are governed by a wave equation with a potential 
concentrated in the vicinity of $r=2M$ in the isotropic 
coordinates used for our examples.  The potential has the effect 
of largely preventing waves from crossing this region.   

\begin{figure}
\begin{center}
\epsfysize=2.7in
\epsfbox{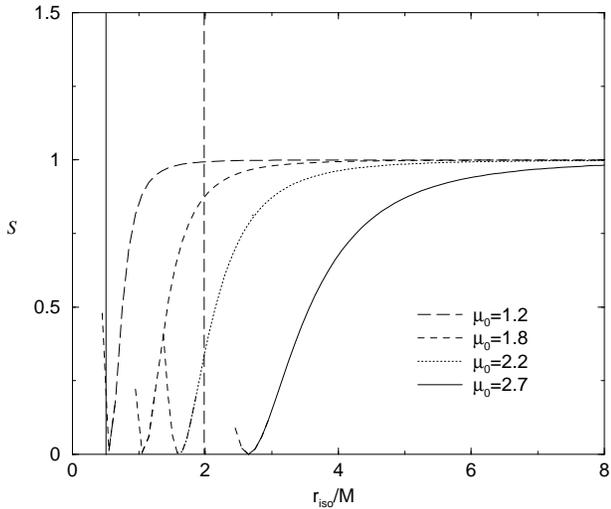}
\end{center}
\caption[The speciality index for Misner data]{The equatorial 
values of ${\cal S}$ shown in this plot allow the interpretation 
that perturbation theory can work for cases of $\mu_0<1.8$ but not higher. 
The ``potential barrier'' near $r_{iso}=2M$ prevents the violation 
of perturbative dynamics inside near the horizon at $r=0.5M$, 
from affecting the radiation.  For larger separations the 
interpretation of a potential barrier at $r_{iso}=2M$ is itself invalidated.}
\label{fig:Sinvs}
\end{figure}

The examples we consider here all correspond to even-parity modes 
implying that ${\cal S}$ is real, so we leave out reference to 
its imaginary component in the following discussion.  
We first consider the case of two initially resting equal-mass 
black holes. This initial configuration is 
represented by the equal-mass time-symmetric Misner datasets 
parameterized by $\mu_0$ as a measure of the initial 
separation. In this case, Schwarzschild black hole perturbation 
theory has been shown to provide a very good estimate of the total 
radiated energy for cases with $\mu_0<1.8$  even though the 
black holes share a common apparent horizon only when $\mu_0 < 1.36$.
Comparisons with numerical calculations\cite{Baker96a} and  
second order calculations\cite{Gleiser96b} have demonstrated that 
the linear perturbation approach overestimates the radiation 
energy by only a factor two up to $\mu_0=1.8$ but beyond that 
the differences grow quickly. Second order perturbations have 
been applied in this case as a useful tool for assessing the 
domain of validity of perturbation theory.
We can obtain similar conclusions 
by applying our speciality index test. 
Fig.\ \ref{fig:Sinvs} shows the initial values of ${\cal S}$ 
along the equator for Misner data, at several 
initial separations. In isotropic coordinates,
 as used in previous studies, the 
horizon of the background black hole is located at $r=0.5M$ and 
location of the perturbative ``potential barrier'' is near $r=2.0M$.
First consider the case, $\mu_0=1.2$.  While there is small region 
near the horizon where the criterion fails, this region is well inside 
the potential barrier which should prevent any local errors in the 
perturbative dynamics from having a significant effect on the outgoing
radiation.  In the marginal $\mu_0=1.8$ case, the error in ${\cal S}$ 
begins to be significant near the potential barrier, and the radiation 
should be somewhat affected.  For larger values of $\mu_0$ the 
violation is significant in the vicinity of the potential barrier 
itself, invalidating the perturbation dynamics.  
The sudden drop in the value of ${\cal S}$ in these cases makes 
our interpretations insensitive to the choice of a 
``factor of two'' cutoff.

\begin{figure}[t]
\begin{center}
\epsfysize=2.7in
\epsfbox{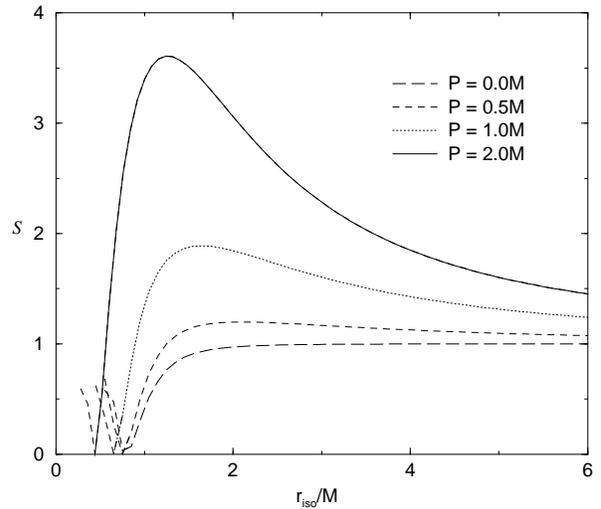}
\end{center}
\caption[Boosted black holes.]{The initial inward linear momentum 
of the black holes results in qualitatively different features in 
the equatorial ${\cal S}$ values.  These curves correspond to 
separations of $\mu_0=1.5$ and suggest that perturbation theory 
should be good below $P=M$.}
\label{fig:SinvsP}
\end{figure}

Another well-studied case is the collision of black holes with 
non-vanishing initial linear momentum $P$.  Here again, 
first order perturbation theory has been very successful, unexpectedly 
providing a good estimate of the radiation energies even for large 
values of $P$\cite{Nicasio99}.
We consider configurations corresponding to a fixed $\mu=1.5$, for 
various values of $P$.   The corresponding initial values 
of ${\cal S}$ are shown in  Fig. \ref{fig:SinvsP}.  
The presence of momentum in the initial slice introduces 
qualitatively different features to the initial values of 
${\cal S}$ exhibiting now a region of ${\cal S}>1$ which falls 
off more slowly at large r.  There is no question in this case 
of the location of this region, but rather the magnitude of the 
violation.  In this case the violation seems to grow quadratically 
with $P$ and reaches a factor of two just after $P=M$.  
Suggesting perturbation theory should be successful up to 
this $P\sim M$.  The prediction of our prescription is consistent with
the accuracy to 10\% for radiation energies shown in Ref. \cite{Nicasio99} 
for $P<M$ but does not explain
the mere doubling of this discrepancy out to $P=3M$.

\begin{figure}
\begin{center}

\epsfysize=2.7in
\epsfbox{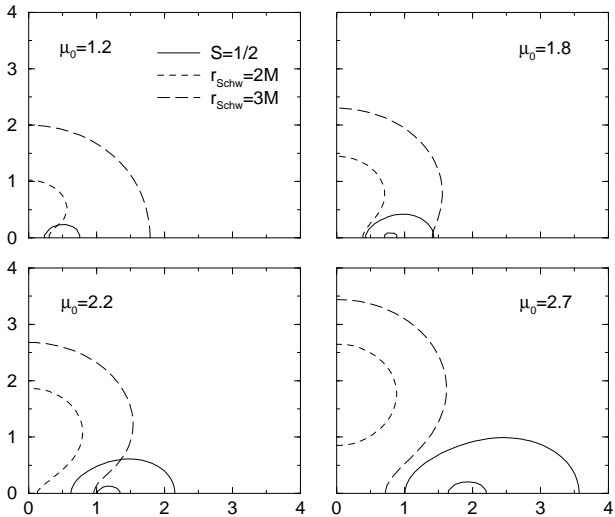}
\end{center}
\caption[Fully invariant analysis]
{Here we reexamine the Misner results, for the 
$\mu_0=1.2, 1.4, 1.8, 2.2$ cases respectively, 
showing the full 2D picture in a quadrant of the $xz$ plane.
The results of the analysis this time provided in terms of 
an invariantly defined background and $r_{Schw}=2M$ horizon and potential 
barrier near $r_{Schw}=3M$.  
The solid curve indicates the ${\cal S}=1/2$ surface.}
\label{fig:all4}
\end{figure}
In the above analysis we have considered the location of the 
${\cal S}\not\approx 1$ regions in relation to the local properties 
of the  ``background'' black hole.  The specification of the 
background metric in this standard treatment is not itself gauge 
independent though.  We can evaluate the applicability of a 
perturbative treatment in a gauge invariant way by utilizing 
a gauge-invariant specification of the background.
A simple way to do this is to specify the background Schwarzschild radial 
coordinate by $r_{Schw}^6=3M^2/I$.  The location of the horizon 
and potential barrier are then found with respect to this 
coordinate.  A two dimensional representation of the results 
for Misner initial data is given in  Fig.\ \ref{fig:all4}.  
These plots show three curves, representing the locations of the 
background horizon, the potential barrier, and the ${\cal S}=0.5$ 
surfaces in a quadrant of the $xz$-plane.  The qualitative features 
observed in the preceeding interpretation of ${\cal S}$ for 
the Misner problem, are reproduced precisely in this more 
complete, fully invariant perspective.    

\section{Discussion}
\label{sec: Discussion}

We have identified a gauge invariant quantity which provides a 
particularly interesting local reduction of the geometric
data in a (numerical) black hole spacetime.  We foresee three
areas of application where $\cal S$ should be a useful quantity:
perturbation studies, which we have discussed in detail, numerical
spacetime interpretation, and numerical code testing.
In the context of perturbation theory we have demonstrated that $\cal S$
provides an invariant criterion for predicting when perturbation theory
might provide a reliable approximation for part of a black hole spacetime. 
This method has the advantage over other estimates such as apparent horizon 
formation that it is genuinely gauge invariant.  In practice it can
serve as simple alternative to second order perturbation theory, but our
prediction does not provide such a direct 
validation of a perturbative calculation.
Knowing that ${\cal S}\sim 1$ does not, for example, 
identify the appropriate background spacetime, a vital
step in any application, but it is very useful predictor of
when perturbation theory may be a useful alternative to numerical simulation
in a generic black hole spacetime.  The quantity $\cal S$ itself is not 
restricted to perturbation studies but should also be useful in the general
interpretation of numerical spacetimes.  As an example, looking at $\cal S$ in 
numerical simulations,  Misner data evolve after a short 
time to qualitatively resemble 
the black hole data with inward momentum discussed above.  
After longer evolutions in 
this family $\cal S$ tends to approach unity in the 
horizon/potential region with
evidence of radiation moving away in both directions.  In this context we also
note that the presence of an {\em isolated horizon}\cite{Ashtekar99a}, 
a recent construct of growing theoretical interest in 
black hole spacetimes,  implies 
locally that ${\cal S}=1.$    Lastly, because $\cal S$ is an 
invariant with often 
predictable behavior, 
it can be very useful in numerical code testing.  In Kerr spacetimes
for example ${\cal S}=1$ 
exactly in any coordinates.  Also in typical cases of Bowen-York 
binary black hole data that 
we have looked at $\cal S$ falls off quickly toward unity 
away from the black holes.  
In light of its simplicity and straightforward significance,
we expect $\cal S$ to become a standard, 
very useful tool for analyzing numerically 
generated spacetimes and interpreting their physical content.

\acknowledgements 
We thank our colleagues at AEI, Richard Price and Jorge Pullin 
for their support and helpful suggestions. This work was supported 
by AEI. M. C. holds a Marie-Curie Fellowship (HPMF-CT-1999-00334).
All our numerical computations have been performed with 
a full 3D code, Cactus, on an 8 GB SGI Origin 2000 with 32 
processors at AEI.
\bibliographystyle{bibtex/prsty}
\bibliography{bibtex/references}
\thebibliography{inv}

\end{document}